\documentclass[twocolumn,pre,english,superscriptaddress]{revtex4-1}
\usepackage{verbatim}
\usepackage{amsmath}
\usepackage{amssymb,color}
\usepackage{graphicx}
\usepackage{babel}
\newcommand{\vecr}{\mathbf{r}}

\begin{document}

\title{Analogies between growing dense active matter and soft driven glasses}


\author{Elsen Tjhung}

\email{elsen.tjhung@durham.ac.uk}

\affiliation{Department of Applied Mathematics and Theoretical Physics, Centre for Mathematical Sciences, University of Cambridge, Wilberforce Road, Cambridge CB3 0WA, United Kingdom}

\affiliation{Department of Physics, Durham University, Science Laboratories, South Road, Durham DH1 3LE, United Kingdom}

\author{Ludovic Berthier}

\affiliation{Laboratoire Charles Coulomb (L2C), Universit\'e de Montpellier, CNRS, 34095 Montpellier, France}

\affiliation{Department of Chemistry, University of Cambridge, Lensfield Road, Cambridge CB2 1EW, United Kingdom}

\begin{abstract}
We develop a minimal model to describe growing dense active matter such as biological tissues, bacterial colonies and biofilms, 
that are driven by a competition between particle division and steric repulsion. 
We provide a detailed numerical analysis of collective and single particle dynamics. 
We show that the microscopic dynamics can be understood as the superposition of an affine radial component due to the global growth, 
and of a more complex non-affine component which displays features typical of driven soft glassy materials, such as aging, compressed exponential decay of time correlation functions, and a crossover from superdiffusive behaviour at short scales to subdiffusive behaviour at larger scales. 
This analogy emerges because particle division at the microscale leads to a global expansion which then plays a role analogous to shear flow in soft driven glasses. We conclude that growing dense active matter and sheared dense suspensions can generically be described by the same underlying physics.
\end{abstract}

\date{\today}

\maketitle

\section{Introduction}

Non-equilibrium physical systems can be divided into two categories: active and externally driven matter. 
In active matter, energy is injected internally and continuously to the system~\cite{Marchetti13,review3}.
Typical examples of active matter include suspensions of microswimmers~\cite{Palacci13,Cates15}, biological tissues \cite{Angelini10,Garcia2015} and 
sub-cellular materials~\cite{Frey10,Prost07}.
On the other hand, in externally driven matter, energy is injected to the system globally or at the boundaries.
Some examples of driven matter are sheared colloidal suspensions~\cite{Larson,Poon07,ballauff} and Rayleigh-Benard convection in fluids~\cite{Faber}. 
One important question in statistical physics is to identify if there is a universality between active and externally driven matter~\cite{Marchetti13,review3}.
For instance in Refs.~\cite{Elsen16,Elsen17,Takeshi16}, it has been suggested that periodically sheared suspensions can be mapped into a particular class of active matter. As another example, dense bacterial suspensions display collective motion that was described using the tools of fluid turbulence~\cite{turbulence}.

We consider {\it growing dense active matter} as a specific type of active system where particles (or mass) are created locally in the system.
Some examples include fast growing and densely packed epithelial tissues~\cite{Angelini10} and bacterial colonies~\cite{colony}.
In the case of tissues, the cells grow up to a certain size before dividing into two daughter cells.
In an open boundary, the dynamics of growing tissues can display some glassy behaviours such as dynamic heterogeneities~\cite{Angelini10}.
On the other hand, under confinement, the dynamics becomes arrested after the density or local pressure reaches a certain value, 
similar to a glass transition~\cite{Hallatschek16,reviewelijah}.
Exploring the role of cell division in dense tissues is relevant to understanding biological processes such as tissue repair and tumor growth \cite{silberzan,silberzan2}.
In numerical simulations, the cells are often approximated as soft spheres without changing much of the physics~\cite{Ranft10,Thirumalai18,Hallatschek19}.
Similarly, some bacteria such as \emph{N. gonorrhoeae} also have a spherical body which changes into a dumbell shape during reproduction~\cite{Welker18},
similar to our numerical model below.
In densely packed colonies, they display liquid/glass-like structure~\cite{Welker18}, although the dynamics has not been much studied experimentally. 
On the other hand, some other bacteria such as \emph{E. coli} and \emph{P. aerigunosa} are rod-shaped instead of spherical~\cite{Poon18,Kragh16}. 
They typically elongate before dividing in the lateral direction. 
This will introduce additional complexity such as liquid-crystalline local order inside the colony~\cite{tsim,Giomi18,Poon18}, which we do not consider further here. 
Collectively, bacteria also exist at large densities in the form of surface-attached biofilms \cite{biofilm1}, whose global structure and microscopic dynamics can now be studied experimentally with high resolution imaging techniques \cite{biofilm2}. Biofilms represent another important research area in biological studies. 

Numerical models devised to study biological systems are often system-specific~\cite{modelbacteria,lardon}. 
It is therefore difficult to identify if there is a unifying physical principle behind all these different examples of growing active matter. 
In this paper, we consider a minimal model of growing dense active matter in two dimensions. 
We consider circular soft repulsive discs which grow and divide in an attempt to capture the competition between two key physical ingredients, 
namely {\it particle division} that increases the density locally, and {\it steric repulsion}.
Furthermore, we regulate the growth and division rates of the particles such that the density inside the growing material remains constant while the total number of particles increases linearly. 
This linear growth is consistent with many experiments of tissue growth~\cite{Montel11,Freyer85,Freyer86}. 
In biofilms, both linear and non-linear scaling such as exponential or algebraic have been reported~\cite{Dockery01,Allen19}.
Other numerical models of tissue growth such as vertex model are also widely studied to take into account confluency~\cite{Manning16,Henkes17,Levine18}.
They are often found to provide very similar dynamics to that of particle-based (or non-confluent) models~\cite{Henkes19}. 
The aim of our work is to isolate the physics of growing and dividing dense active matter without any additional system-specific interactions such as confluency~\cite{Manning16}, cell-cell adhesion~\cite{Ranft10} or non-sphericity of the particles~\cite{Giomi18}, 
and removing all other sources of motion such as thermal fluctuations and self-propulsion. 
In real biological systems, all these factors can of course coexist. 

It has been argued that cell division can fluidize active tissues because particle division necessarily produces a local rearrangement~\cite{joanny}. 
So, when a sufficient number of particle divisions has taken place, the entire system has been structurally reorganised, and this allows the material to flow~\cite{joanny,silke}. 
We argue below that this simple intuition does not account for the dynamic behaviour seen in growing active matter, where the global expansion itself plays a key physical role in the fluidisation of the material.

In recent work, it was shown that growing tumors display dynamic behaviour reminiscent of supercooled liquids approaching a glass transition~\cite{Jimenez-Valencia15,Thirumalai18}. 
We shall similarly propose an analogy with glassy materials~\cite{Berthier11}, but will provide evidence that growing active matter actually resembles glassy materials that are externally driven at large scale, such as sheared colloidal suspensions~\cite{Poon07,ballauff}. 
To reach this conclusion we show that the single particle dynamics should be decomposed into two distinct components. 
First, particles move radially in the growing material as a response to the global expansion. 
These affine displacements are slaved to the radial growth of the colony and can overshadow the second, more complex non-affine component of the displacement. 
Removing affine displacements is standard practice for sheared glasses~\cite{Falk98}, 
but rarely performed in active matter and biological physics literature until very recently~\cite{Prakash19}.

After subtracting the affine particle displacements, we show that the non-affine dynamics of the particles are spatially heterogenous and display aging behaviour. 
In addition, we find that the time correlation functions display compressed exponential decay at short length scales, 
a signature found in many soft glassy materials~\cite{Cipelletti00,ramos05,bob,ruta}. 
The mean squared displacements show a crossover from superdiffusive behaviour at short time scales to subdiffusive behaviour at long time scales. 
Overall, we conclude that this characteristic aging dynamics is not directly caused by the individual local division events, 
but is instead controlled by the global radial growth rate of the colony which indeed results from particle division. 
The radial growth rate plays therefore the same role as the global shear rate in sheared dense suspensions~\cite{Poon07,ballauff}. 
This suggests that both growing active matter and externally driven soft glassy matter are described by the same underlying physics.

Our paper is organised as follows. 
In Sec.~\ref{model} we introduce the model and describe its macrocopic behaviour.
In Sec.~\ref{aging} we characterise its aging microscopic dynamics. 
In Sec.~\ref{dynhet} we demonstrate the heterogeneous character of the dynamics and show that it is driven by the global expansion of the material.
In Sec.~\ref{1D} we consider a quasi 1-dimensional geometry to show that our result remains robust with different geometries.
In Sec.~\ref{conclusion} we discuss our results and offer some perspectives for future work.

\section{Model and macroscopic behaviour}

\label{model}

\begin{figure}
\begin{centering}
\includegraphics[width=1.0\columnwidth]{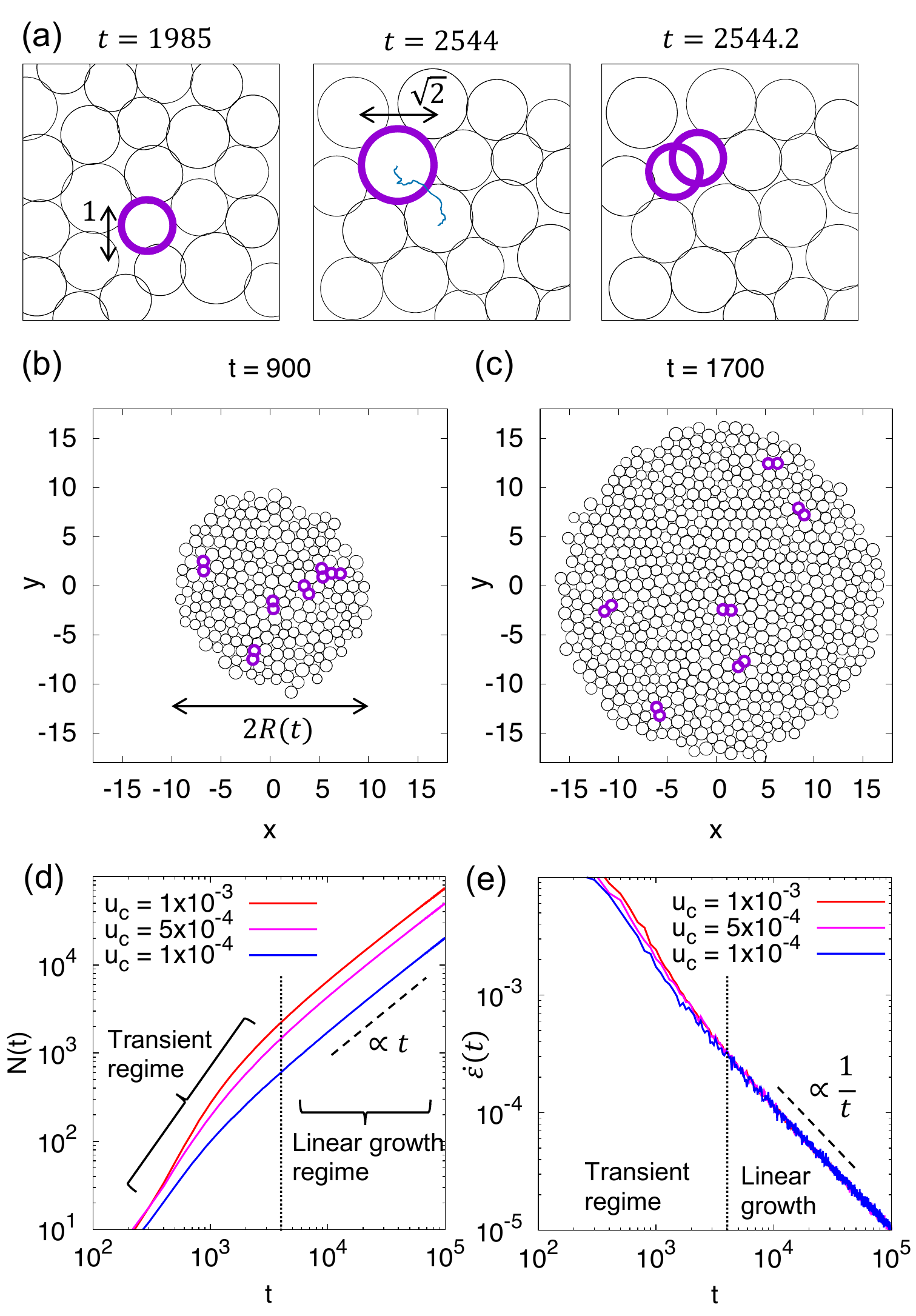}
\par\end{centering}
\caption{(a) Model: each particle grows with a rate that depends on the total energy density. When a given diameter $\sigma_i(t)$ reaches the critical size $\sqrt{2}$, particle $i$ divides into two particles with equal diameters $\sigma=1$. 
(b,c) Snapshots of the growing colony at different times $t$. Highlighted particles have recently divided within time interval $[t,t-10]$.
(d) The total number of particles $N(t)$ grows linearly with $t$ after a transient regime. 
(e) Consequently, the global strain rate $\dot\varepsilon(t)=\dot{R}/R$, decays as $1/t$ at large times. \label{fig:model}}
\end{figure}

\subsection{Model}

We consider $N(t)$ circular particles in an infinite two-dimensional space.
The number of particles, $N(t)$, is not conserved but grows with time $t$, due to the dynamics which we introduce below.

Let us denote $\vecr_i(t)$ and $\sigma_i(t)$ as the position and diameter of particle $i$ at time $t$ respectively.
The diameter of the particles varies from $\sigma$ to $\sqrt{2}\sigma$.
The interaction between the particles is approximated by a two-body repulsive harmonic potential,
\begin{equation}
V(r_{ij},\sigma_{ij}) = 
\begin{cases} 
\frac{\epsilon}{2} \left(\frac{\sigma_i^2+\sigma_j^2}{2}\right) \left(1- \frac{r_{ij}}{\sigma_{ij}} \right)^2, & \,\, r_{ij}\le\sigma_{ij}, \\
0, & \,\, r_{ij}>\sigma_{ij}, \label{eq:potential}
\end{cases}
\end{equation}
where $r_{ij}=|\vecr_i-\vecr_j|$ and $\sigma_{ij}=(\sigma_i+\sigma_j)/2$. 
This potential has been used before in the context of foams~\cite{Durian95}, soft tissues~\cite{Ranft10,Thirumalai18}, and biofilms~\cite{modelbacteria,Farrell13,Allen19}.
Note that in some models of tissues~\cite{Ranft10,Thirumalai18}, 
there is also a short-range attractive potential to mimic cell-cell adhesion, which we do not include here, because it is not needed to maintain the tissue integrity. Another reason is that we aim to discover some generic physical principles in all growing active matter, which also includes bacterial colonies where adhesion is not necessarily present.
In Eq.~(\ref{eq:potential}), $\epsilon$ is the interaction strength (in units of energy per unit area). The prefactor $(\sigma_i^2 + \sigma_j^2)/2$ in (\ref{eq:potential}) is chosen so that the total energy, $\frac{1}{2}\sum_{ij}V(r_{ij},\sigma_{ij})$,
scales linearly with the total area of the particles $\sum_i\pi\sigma_i^2/4$.

We also assume that thermal fluctuations are unimportant, thus, the dynamics of the particles are given by the overdamped equations:
\begin{equation}
\xi\frac{d\vecr_i}{dt} = -\sum_{j=1}^{N(t)}\frac{\partial V(r_{ij},\sigma_{ij})}{\partial\vecr_i},
\label{eqmotion}
\end{equation}
where $\xi$ is the friction coefficient describing the interaction with the substrate. Together with $\epsilon$, this defines the microscopic time scale  for energy dissipation, namely $\tau_d=\xi \sigma^2/\epsilon$. Physically, this represents the typical time for two initially overlapping particles to move away from each other. In all simulations presented below, we shall work in reduced units where $\sigma=\tau_d=1$.

The dynamic equations of motion in Eq.~(\ref{eqmotion}) ensure that repulsion due to particle crowding is described in the simplest possible manner. In the absence of growth and division, the system would quickly arrive at rest in a state without any particle overlap since the system has open boundaries. 

To continuously drive the dynamics and make the system active, we now introduce the growth and division dynamics for individual particles.    
Let us define the total energy density at time $t$:
\begin{equation}
u(t) = \frac{4}{N\pi\sigma^2}\sum_{i\neq j}V(r_{ij},\sigma_{ij}).
\end{equation}
We assume that each particle's diameter, $\sigma_i(t)$, increases with a growth rate which depends on $u(t)$:
\begin{equation}
\frac{d\sigma_i}{dt} = 
\begin{cases}
\alpha_i \left(\frac{u_c-u(t)}{u_c} \right), &\,\, u(t) \le u_c, \\
		0,					   & \,\, u(t) > u_c, \label{eq:rate}
\end{cases}
\end{equation}
where $\alpha_i$ is a constant, which is different for each particle $i$.
In practice, it is drawn from a uniform random distribution $[0,\alpha_{\text{max}}]$. 
The value of $\alpha_{\text{max}}$ does not affect the main result in this paper, 
as long as it is chosen to be much smaller than the dissipative timescale $\tau_d=1$, in order to be close to a quasi-static limit with only very small particle overlaps.
We choose the characteristic growth rate $\alpha_{\text{max}}=0.01$.
For the division we introduce the following rule. 
When the particle diameter reaches a maximal size $\sqrt{2}$, the particle divides into two particles with equal diameters $\sigma=1$, as illustrated in Fig.~\ref{fig:model}(a). Images also show that this choice of parameter does not lead to any particular spatial correlation between the diameters of the particles which are constantly mixed during the growth of the system.

In Eq.~(\ref{eq:rate}), 
$u_c$ is the energy density threshold, such that after some initial transient regime, the global energy density will saturate around $u(t)\simeq u_c$.
Equivalently, the growth rate (\ref{eq:rate}) can also be defined \emph{via} the isotropic Kirkwood stress, with a corresponding pressure threshold~\cite{Thirumalai18}. Consequently, the number of particles $N(t)$ increases linearly (instead of exponentially if we took a constant growth rate) with time, 
except in the initial transient regime at $t\lessapprox3000$, see Fig.~\ref{fig:model}(d).
This linear growth is consistent to many experiments in tumour tissue growths~\cite{Montel11,Freyer85,Freyer86}.
In a two-rate model~\cite{Radszuweit09}, the particles on the perimeter usually divide at a quicker rate compared to particles in the bulk.
In our simplified model, we assume homogenous division rate for all particles in the colony.

\subsection{Macroscopic behaviour}

We start all simulations with a single particle with unit diameter at the origin at time $t=0$. We then let the particle grow, divide, and so on.
Figure~\ref{fig:model}(b,c) shows the snapshots of this growing colony at two different times $t$. The system remains roughly circular, characterized by the radius $R(t)$. 
After some initial transient regime $t\lessapprox3000$, 
the number of particles grows linearly with time. 
We call this the linear growth regime, see Fig.~\ref{fig:model}(d).
In the linear growth regime, the area fraction inside the colony, $\phi_c$, is roughly uniform in space and constant in time.
The value of $\phi_c$ depends on the chosen critical energy density $u_c$.
For example, for $u_c=0.001$, $\phi_c\simeq0.92$, which is higher than the jamming density $\phi_J\simeq0.84$ below which particles would no longer overlap.
In the following, we fix  $u_c=0.001$, but we have also verified that the results in this paper do not change if one uses instead $u_c=0.0001$ 
(corresponding to $\phi_c\simeq0.87$).

An important physical quantity resulting from the linear growth of the system is the global radial growth rate defined as
\begin{equation}
\dot{\varepsilon}(t) = \frac{\dot{R}(t)}{R(t)} ,
\label{eq:varepsilon}
\end{equation}
where $R(t)$ is the radius of the colony at time $t$, see Fig.~\ref{fig:model}(b).
In the linear growth regime, the area fraction inside the colony is roughly constant and $N(t)\propto t$,
thus, it follows that $\dot{\varepsilon}(t)\propto1/t$, see Fig.~\ref{fig:model}(e). We shall be mostly concerned with the dynamics of the material in the linear growth regime at large enough times. 

The expansion of the colony is driven by local division events, and thus the system is a genuine active system because energy injection occurs at the particle scale and no other {\it external} driving forces are present, in particular at larger scale. However, we shall argue below that the dynamics of the particles can be described solely by using the global variable $\dot{\varepsilon}(t)$, which can be seen as a global mechanical forcing acting on the colony, induced by the particle activity at small scale. Therefore the dynamics of growing active matter at high density is poised to resemble the one of other externally driven dense systems~\cite{SGR,BBK}  such as sheared dense suspensions, where the global driving force is directly applied at large scale by an operator~\cite{Poon07,ballauff}. 

Finally, since the growth rate $\dot{\varepsilon}$ is decaying with time as $1/t$, we must take into account that the effective global driving force depends on time, and thus we must expect that the dynamics is going to display aging phenomena~\cite{agingstruik}. Because  the driving force $\dot{\varepsilon}$ decreases with time, we expect the system to become slower at large times, again in full analogy with aging glassy materials. When measuring time correlation functions, it will therefore be useful to introduce notations to aging materials. We shall measure the dynamics between two times $t_w$ and $t_w + \Delta t$, where $t_w$ is the waiting time since the simulations started, and $\Delta t$ is the time interval over which dynamics is analysed. In an aging material, time translational invariance is lost, and dynamics does not uniquely depend on $\Delta t$ but also on $t_w$. It is often found that the scaled variable $\Delta t / t_w^{\mu}$ collapses time correlation functions at large times~\cite{agingstruik,agingSG}. The exponent $\mu \approx 1$ would correspond a simple aging scaling, whereas $\mu < 1$ has been called sub-aging, a behaviour reported in many different types of disordered materials~\cite{agingstruik,agingSG} and model systems~\cite{BB02,BY04}.  

\section{Microscopic aging dynamics}

\label{aging}

\subsection{Affine and non-affine displacements}

\begin{figure}
\begin{centering}
\includegraphics[width=1.\columnwidth]{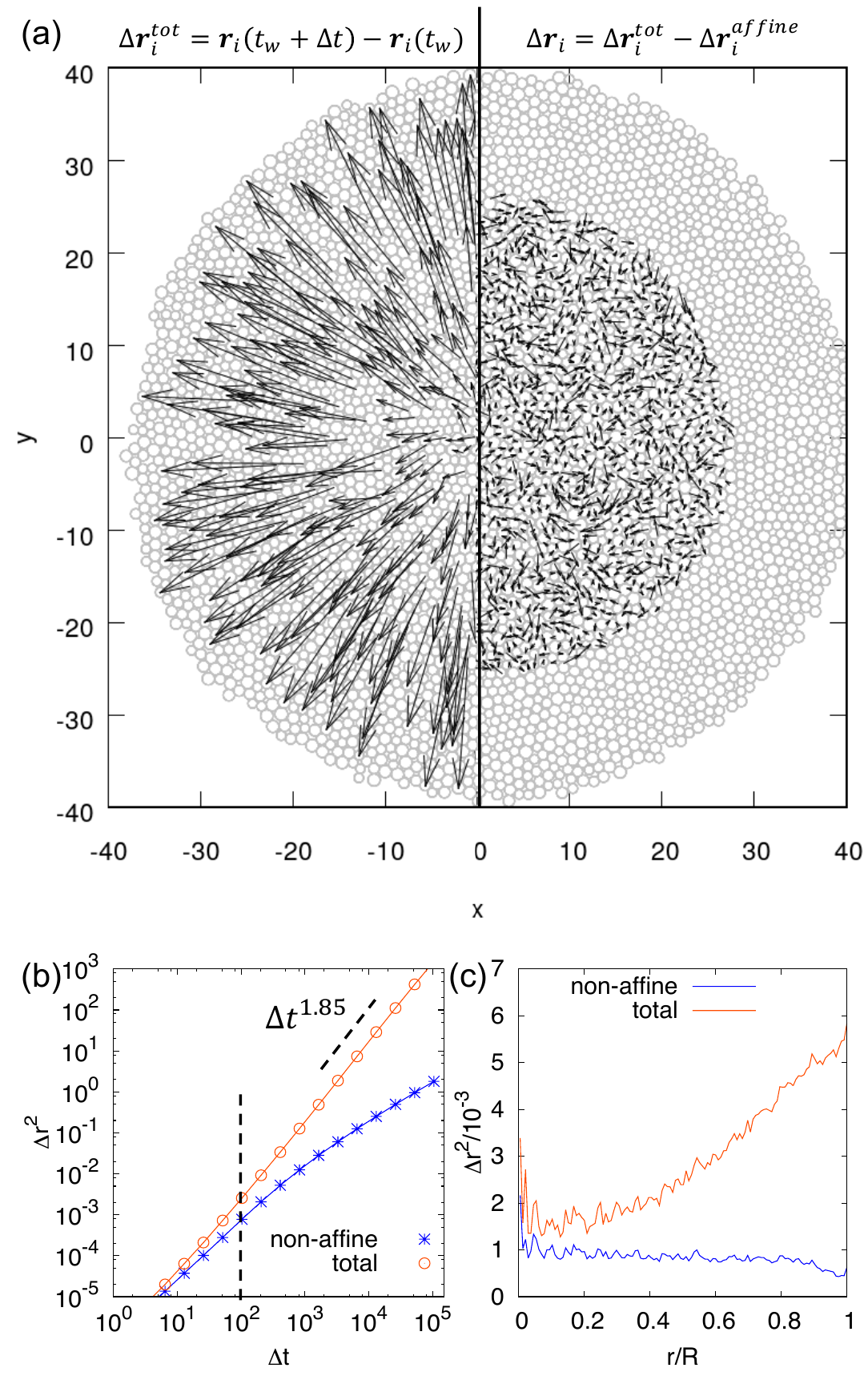}
\par\end{centering}
\caption{
(a) Left: Total displacements of the particles from time $t_w$ to $t_w+\Delta t$ consist of mainly large affine displacements in the radial direction.
Right: After subtracting the affine component,
the non-affine displacements of the particles, $\Delta\vecr_i$, reveal a collective, aging, and heteregeneous dynamics. 
Here, $t_w=3276.8$ and $\Delta t=3276.8$.
(b) Comparison of the total mean squared displacement (MSD) and the non-affine MSD for $t_w=209715.6$.
(c) The non-affine MSD is fairly homogeneous but the total MSD increases linearly with the radial distance from the centre.
Here, $t_w=209715.6$ and $\Delta t=25.6$.
\label{fig:non-affine}}
\end{figure}

We now investigate the aging microsopic dynamics, focusing mainly on the linear growth regime where $N(t)\propto t$.
In this regime, the density inside the colony is roughly constant and particles are created uniformly anywhere inside the colony.
Thus, we expect an \emph{affine velocity field} in the radial direction for every particle $i$:
\begin{equation}
\left(\frac{d\vecr_i}{dt}\right)^\text{aff} = \frac{1}{2} \frac{\dot{N}(t)}{N(t)}  \vecr_i(t) \simeq \dot{\varepsilon}(t) \vecr_i(t),
\label{eq:aff}
\end{equation}
where we also assume the interface of the colony remains circular (see Fig.~\ref{fig:model}(b,c)).
The affine radial velocity of the particle is proportional to the radial growth rate $\dot{\varepsilon}$ at that particular time $t$ and to the distance of the particle from the origin.
(Note that the colony is centred around the origin of the coordinate system.)
Consequently, if we measure the total mean squared displacement (MSD) of the particles from time $t_w$ to $t_w+\Delta t$, these displacements should be influenced by the affine growth of the system. In the limit where affine displacements dominate, superdiffusion would be observed at long times even if particles do not actually relax the structure of the system. Indeed, previous numerical simulations~\cite{Thirumalai18} and experimental studies~\cite{Jimenez-Valencia15} reported superdiffusive particle motion.
This superdiffusive behaviour in the total MSD is perhaps unsurprising because of the finite average velocity in the radial direction,
giving rise to large affine displacements, see Fig.~\ref{fig:non-affine}(a).
However, the total MSD is not necessarily ballistic either, because the radial growth rate is actually decaying as $1/t$ and thus the affine velocity also decays with time.

In order to uncover the non-trivial dynamics in these growing dense active matter, we first need to subtract these affine contributions from the total displacements of the particles.
The \emph{affine displacement} of particle $i$ from time $t_w$ to $t_w+\Delta t$ is the time integral of the affine velocity:
\begin{equation}
\Delta\vecr_i^\text{aff} = \int_{t_w}^{t_w+\Delta t} \dot{\varepsilon}(t) \vecr_i(t)\,dt. \label{eq:aff}
\end{equation}
We define the \emph{non-affine displacement} of particle $i$ to be the difference between the total displacement and the affine displacement: 
\begin{equation}
\Delta\vecr_i (t_w,\Delta t)= \Delta\vecr_i^\text{tot} - \Delta\vecr_i^\text{aff},
\end{equation}
where the total particle's displacement is simply:
\begin{equation}
\Delta\vecr_i^\text{tot} (t_w,\Delta)= \vecr_i(t_w+\Delta t)-\vecr_i(t_w).
\label{eq:rtot}
\end{equation}
This decompostion is illustrated in Fig.~\ref{fig:non-affine}(a).
Note that during this time interval $[t_w,t_w+\Delta t]$, new particles are created due to division events, but we do not track these new particles. They of course will be tracked if we start a new measurement at a later time $t_w$. 

In previous literature~\cite{Thirumalai18,Jimenez-Valencia15}, 
the MSD is computed based on the total particle's displacement in Eq.~(\ref{eq:rtot}), $\left<|\Delta\vecr_i^\text{tot}|^2\right>$, without subtracting the affine components. Instead we subtract the affine components to define the non-affine MSD as 
\begin{equation}
\Delta^2 r (t_w, \Delta t) = \langle | \Delta \vecr_i (t_w, \Delta t) |^2 
\rangle.
\end{equation}
In Fig.~\ref{fig:non-affine}(b) we illustrate that the total and affine components of the motion behave similarly at very short time intervals $\Delta t$, but strongly deviate from one another at larger times. Whereas a strong superdiffusion is observed from the total displacements, we observe a non-trivial transition between superdiffusive behaviour at short times to subdiffusion at longer times from the non-affine MSD. 

\begin{figure*}
\begin{centering}
\includegraphics[width=0.9\textwidth]{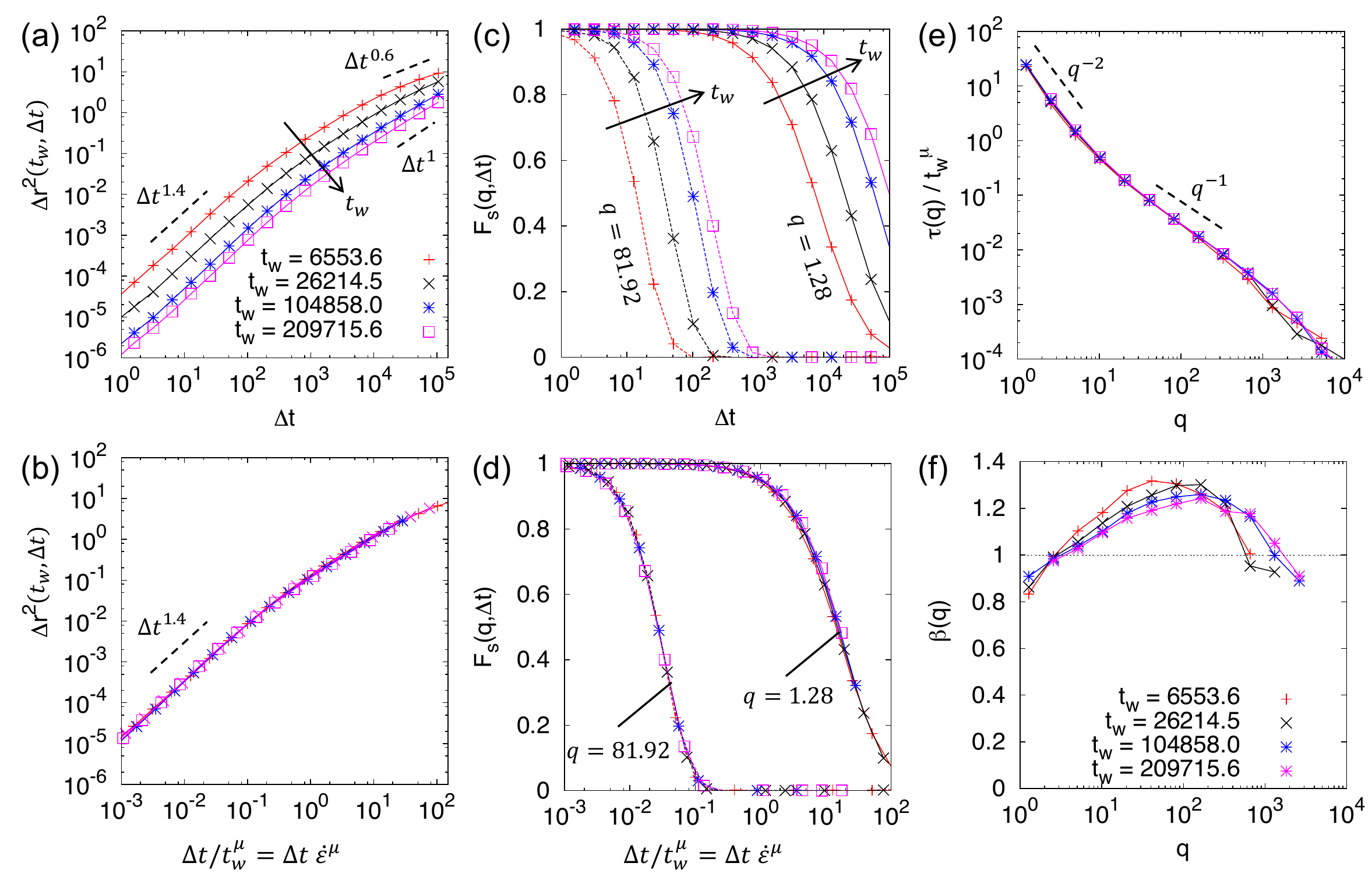}
\par\end{centering}
\caption{
(a) Non-affine mean squared displacements as a function of delay time $\Delta t$ and various waiting times $t_w$ display a crossover from superdiffusive to subdiffusive behaviour.
(b) The scaled variable $\Delta t /t_w^{\mu}$ with $\mu \approx 0.7$ rescales the MSD onto a mastercurve. 
(c) Intermediate scattering function as a function of delay time $\Delta t$ and various waiting times $t_w$ for wavevectors $q=1.28$ (solid lines) 
and $q=81.92$ (dashed lines).
(d) Rescaled intermediate scattering functions with the same 
variable $\Delta t /t_w^{\mu}$.
(e) Collapsed $q$-dependent relaxation timescales $\tau(q) / t_w^{\mu}$, displaying a crossover from ballistic $q^{-1}$ at high $q$, to approximately diffusive $q^{-2}$ at low $q$.
(f) Compressed exponential decay, $\beta(q)>1$, is observed at all $t_w$ over a broad range of wavevectors.
\label{fig:aging}}
\end{figure*}

In Fig.~\ref{fig:non-affine}(c) we spatially resolve the MSD as a function of the radial position within the circular system (the system is by construction rotationnally invariant). As expected from Eq.~(\ref{eq:aff}), we find that the affine component of the displacement increases roughly linearly with the radial distance from the center, such that particles at the boundaries are advected by the radial growth much faster than the ones in the center. This effect should largely explain the strong radial dependence of the total MSD recently reported in Ref.~\cite{thirumalai2019}.  By contrast, we find that the non-affine displacement is fairly homogeneous throughout the system, except for a thin particle layer very near the open boundary. 

In what follows, we will only consider the non-affine displacements of the particles to study the microscopic aging dynamics of the growing active system, and this dynamics will be averaged over the entire system.

\subsection{Sub-aging dynamics}

We start by discussing the aging behaviour of the MSD defined from the non-affine displacements. The numerical data are shown in Fig.~\ref{fig:aging}(a) as a function of delay time $\Delta t$ for different waiting times $t_w$. 

We first observe that for any given $t_w$ the MSD shows a crossover from superdiffusive behaviour at small delay times and small displacements, to a subdiffusive behaviour at long times and large displacements. 
The superdiffusive regime is characterised by an exponent $\Delta^2 r \sim \Delta t^{\alpha}$ with $\alpha \approx 1.4$, which is intermediate between diffusive ($\alpha=1$) and ballistic ($\alpha=2$). The reason for this intermediate behaviour will become clear in Sec.~\ref{dynhet} below. On the other hand, for large $\Delta t$, the non-affine MSD becomes subdiffusive with exponent ranging from $\alpha \approx 0.6$ to $\alpha \approx 1$, see Fig.~\ref{fig:aging}(a). Note that in order to determine the asymptotic value of the exponent in the long time limit $\Delta t / t_w \rightarrow\infty$, much longer simulation would be required. 

The second observation from the MSD data in Fig.~\ref{fig:aging}(a) is the fact that the curves for different times $t_w$ do not superimpose. Instead, the dynamics becomes slower as the age $t_w$ of the system increases. Therefore, the microscopic dynamics of the growing material is not time-translational invariant and slows down with $t_w$: the system is aging. 
This is expected as the growth rate $\dot{\varepsilon}$ (which acts as the driving force) also becomes slower with increasing time.

A third noticeable aspect of the MSD data is the absence of a plateau regime at intermediate timescales. This plateau would represent the well-known caging dynamics often observed in glassy materials approaching a glass transition~\cite{Berthier11}. No such plateau can be expected in our simulations despite the fact that the system is very crowded, because we did not include any thermal fluctuations in the equations of motion Eq.~(\ref{eqmotion}). In fully athermal systems, typical MSD indeed do not display any signature of caging dynamics~\cite{atsushi}, and when sheared they also display a similar crossover from superdiffusive to diffusive~\cite{atsushi,klaus}. We expect that a small amount of temperature in our system would introduce vibrational short time dynamics. This would reveal caging dynamics, but this could also potentially obscure the short time superdiffusive behaviour reported in Fig.~\ref{fig:aging}(a).   

It turns out that we can collapse the MSD for different $t_w$ into a single universal function by rescaling the time delay $\Delta t$ in the horizontal axis by an algebraic function of $t_w$, introducing the rescaled variable $\Delta t / t_w^{\mu}$, where the exponent $\mu \approx 0.7$ is therefore a sub-aging exponent characterizing the aging dynamics, see Fig.~\ref{fig:aging}(b). Polymers and spin glasses are often characterised by a very similar sub-aging exponent~\cite{agingstruik,agingSG,BB02,BY04}. 

\subsection{Compressed exponential decay of time correlations}

Another way to probe the relaxation dynamics in the system is to measure the intermediate scattering function,
\begin{equation}
F_s(q, t_w, \Delta t) = \left< \frac{1}{N(t_w)}\sum_{i=1}^{N(t_w)} e^{i\mathbf{q}\cdot\Delta\vecr_i}\right>, \label{eq:Fs}
\end{equation}
where $\Delta\vecr_i (t_w,\Delta t)$ is the non-affine displacement of particle $i$ within time interval $[t_w,t_w+\Delta t]$, and $N(t_w)$ is the total number of particles at the time $t_w$ where we start the measurement. Since the system is isotropic, $F_s$ only depends on the modulus of the wavevector $q=|\mathbf{q}|$. Physically, $F_s$ is related to the fraction of particles which have moved a distance larger than $2\pi/q$ during time interval $[t_w,t_w+\Delta t]$. It contains similar information to the MSD, but the dynamics is now resolved in space, because large wavevectors probe the dynamics at short distances, whereas small wavevectors probe large displacements. In addition, the intermediate scattering function is accessible to light scattering experiments, whereas the MSD is more suitable to experiments where particle positions can be resolved, for instance using real space imaging techniques. 

Figure~\ref{fig:aging}(c) shows $F_s(q,t_w,\Delta t)$ as a function of delay time $\Delta t$ for different waiting times $t_w$ and 
two very different wavevectors $q=1.28$ (solid lines) and $q=81.92$ (dashed lines). 
For small $\Delta t$, the displacements are small and thus $F_s \simeq 1$. As $\Delta t$ increases, the configuration becomes less correlated with the initial condition at $\Delta t=0$ and thus $F_s$ decays to zero. Similarly to the MSD, we observe aging behaviour in $F_s(q,\Delta t)$ because $F_s$ decays slower when the waiting time $t_w$ increases. Likewise, $F_s(q,t_w, \Delta t)$ for different waiting times can be collapsed into a single mastercurve by rescaling the delay time using the variable $\Delta t / t_w^{\mu}$, as shown in Fig.~\ref{fig:aging}(d). Remarkably, the same sub-aging exponent as for the MSD, $\mu \approx 0.7$, can be used to collapse the self-intermediate scattering function at all wavevectors. 

\begin{figure*}
\begin{centering}
\includegraphics[width=1.\textwidth]{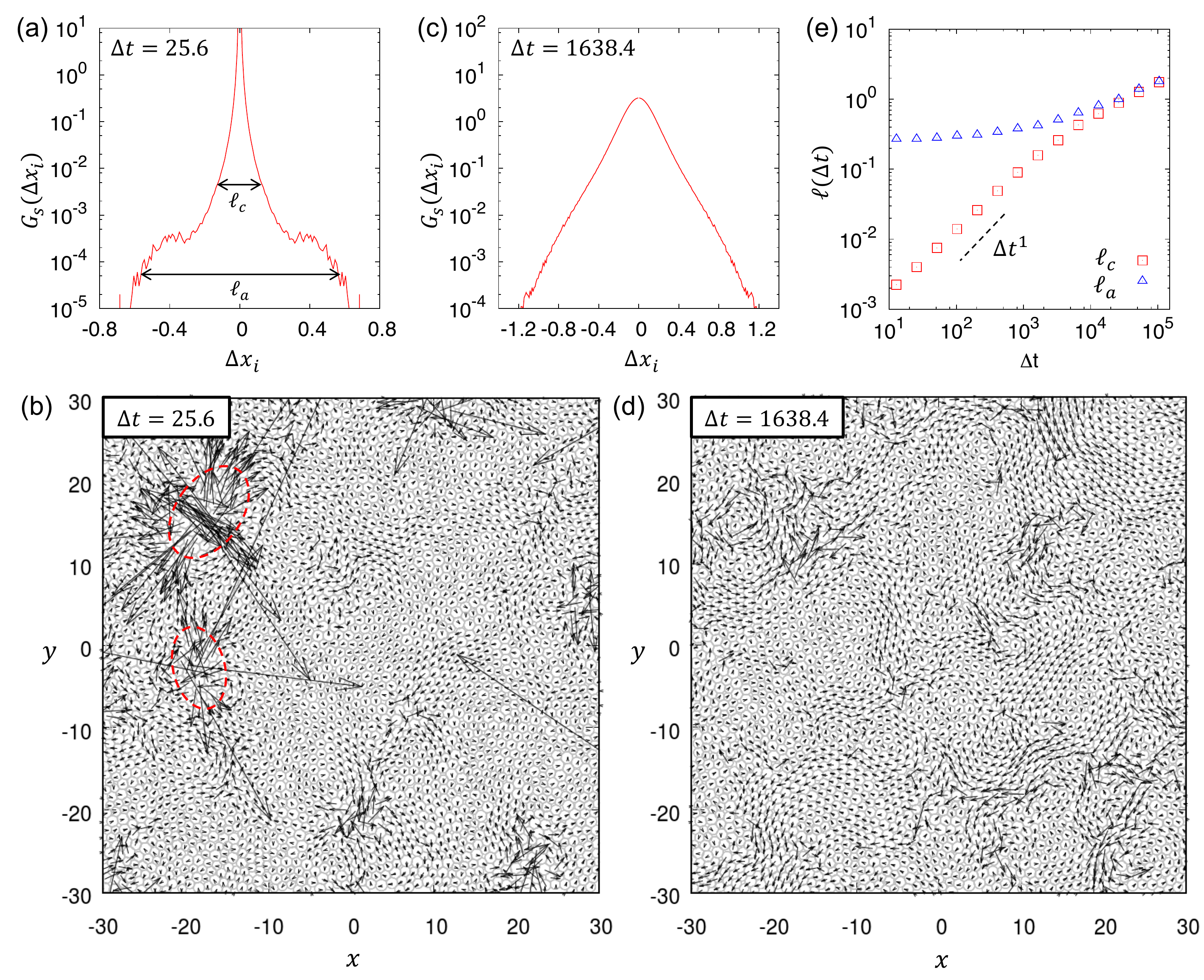}
\par\end{centering}
\caption{
(a,c) Van Hove particle distribution, Eq.~(\ref{eq:gs}) for $t_w=104858.0$. 
In (a), there is a clear separation of length scales into $\ell_a$ and $\ell_c$.
(b,d) Non-affine displacement fields $\Delta\vecr_i$ corresponding to (a,c), the field is respectively magnified by a factor of $120$ and $4$ for better visibility. Division events are highlighted with dashed circles in (b).
(e) Time evolution of the lengthscales $\ell_a$ and $\ell_c$.
\label{fig:heterogeneity}}
\end{figure*}

Empirically, we find that $F_s$ can be well fitted using the following functional form:
\begin{equation}
F_s(q,t_w, \Delta t) \simeq \exp \left[ - \left( 
\frac{\Delta t}{\tau}\right)^{\beta} \right],
\end{equation}
where $\tau(q,t_w)$ is a relaxation timescale and $\beta(q,t_w)$ is called the  stretching (when $\beta < 1$) or compression (when $\beta>1$) exponent.

Given that the global dynamics slows down with the variable $t_w^\mu$, it is natural to consider the rescaled relaxation times, $\tau(q,t_w) / t_w^{\mu}$, as shown in Fig.~\ref{fig:aging}(e). 
As expected this rescaling collapses the data on a unique $q$ dependent function which describes the dynamics of the system as a function of the probed lengthscale. 
At large $q$ (short distances), we find that $\tau(q) \approx q^{-1}$ which corresponds to ballistic dynamics (where displacement is proportional to time). 
This behaviour is only qualitatively consistent with the short-time behaviour displayed by the MSD which was not fully ballistic but only superdiffusive with an exponent $\alpha \approx 1.4$ in Fig.~\ref{fig:aging}(a). 
The reason for these seemingly distinct behaviours will be elucidated in Sec.~\ref{dynhet}. 
On the other hand, the small $q$ regime of $\tau(q,t_w)$ appears to be fully diffusive.

Finally, we show in Fig.~\ref{fig:aging}(f) the $q$-dependence of the exponent $\beta(q,t_w)$ characterizing the functional form of the time decay of the correlation function. We find that $\beta(q,t_w)$ is a non-monotonic function of the wavevector, being close to unity at both large and small distances, which corresponds to nearly exponential decay. However, at intermediate lengthscales, we find that $\beta>1$, which corresponds to a compressed exponential decay. 
Such compressed exponential decay has often been reported in soft glassy systems~\cite{Cipelletti00,ramos05,bob,ruta,pinaki}, in conjunction with ballistic regime $\tau(q) \sim q^{-1}$. 
Therefore, growing active matter represents one more example of a non-equilibrium soft matter system displaying anomalous aging dynamics. 

\section{Spatially heterogeneous dynamics}

\label{dynhet}

\subsection{Coexistence of two typical displacement scales}

We noticed above that the MSD and the self-intermediate scattering function were giving somewhat different indications regarding the ballistic and superdiffusive character of particle motion at short times. Given that both quantities are based on the statistics of single particle displacements, they can only provide distinct quantitative information in the case where the underlying distribution of particle displacements is strongly non-Gaussian, thereby revealing the existence of a strong dynamic heterogeneity~\cite{bookDH,physics}.
 
To investigate this issue further, we measure the distribution of single particle displacements (also called the van Hove function)
\begin{equation}
G_s(\Delta x,t_w,\Delta t) = \langle \delta(\Delta x - |\Delta x_i(t_w,\Delta t)|) \rangle,  
\label{eq:gs}
\end{equation}
where $\Delta x_i$ is the projection of $\Delta \vecr_i$ onto the $x$-axis (isotropy implies that we can average over both $x$ and $y$ directions). We show some representative distributions in Figs.~\ref{fig:heterogeneity}(a,c), whereas Figs.~\ref{fig:heterogeneity}(b,d) show the corresponding non-affine displacement fields $\Delta\vecr_i$ in real space. The waiting time is fixed at $t_w=104858.0$, while we look at two different delay times $\Delta t=25.6$ and $\Delta t=1638.4$. For longer time delays, nearly Gaussian distributions are revealed and these distributions and the corresponding snapshots are thus not very interesting.

As can be seen from Fig.~\ref{fig:heterogeneity}(a), for short delay time $\Delta t=25.6$, the probability distribution $G_s$ shows a clear separation of length scales between a narrow core of slowly moving particles, and large tails of fast moving particles. 
The coexistence of (many) slow and (few) fast particles inside a single sample leading to non-Gaussian van Hove functions is a hallmark of dynamic heterogeneity in glassy materials~\cite{bookDH}. 
To quantify the coexistence of slow and fast particles, we find it empirically convenient to fit the two parts of the distribution $G_s(\Delta x, t_w, \Delta t)$ as the sum of two Gaussian distributions characterised by two distinct spatial extensions, $\ell_c (t_w,\Delta)$ and $\ell_a (t_w,\Delta t)$, as sketched in Fig.~\ref{fig:heterogeneity}(a).  
Note that in Fig.~\ref{fig:heterogeneity}(a), $G_s$ has both a Gaussian tail and a Gaussian peak. 
This contrasts slightly to other active/driven disordered materials, which usually have an exponential tail~\cite{Elsen16,Kob}.
 
The larger length scale, $\ell_a$ is associated with irreversible plastic events in the system which can be directly caused by division events (see the 
red circles in the corresponding displacement fields in Fig.~\ref{fig:heterogeneity}(b)), but not always. There are also many plastic events leading to large particle displacements in locations where no division has occurred during $t_w$ and $t_w+\Delta t$. The smaller length scale, $\ell_c$, is instead associated with the collective motion of the particles that need to respond to the localised plastic events. Similar observations of localised plastic events leading to large scale collective motion are routinely  made in the field of sheared glasses~\cite{bookDH}, where the localised events are called shear transformation zones~\cite{Falk98}, and the collective motion results from the stress redistribution carried by the elastic medium~\cite{picard}. 

At longer delay time $\Delta t=1638.4$ (see Fig.~\ref{fig:heterogeneity}(c)), both $\ell_c$ and $\ell_a$ tend to come closer. This is confirmed in Fig.~\ref{fig:heterogeneity}(e), which describes the $\Delta t$ evolution of the two length scales for a fixed $t_w$. Interestingly, we observe that 
the small length scale $\ell_c$ is ballistic at short delay time, namely $\ell_c \propto \Delta t$. By contrast, the larger length scale $\ell_a$ is roughly constant in this regime, and depend neither on $\Delta t$ nor on $t_w$. Therefore, $\ell_a$ quantifies the typical particle displacement in localised plastic events and it is always of the order of a fraction of the particle diameter. In the long $\Delta t$ limit, both $\ell_c$ and $\ell_a$ seem to converge together, see Fig.~\ref{fig:heterogeneity}(e), and the distinction between large and small displacements becomes irrelevant as the van Hove function becomes a Gaussian in the limit $\Delta t\rightarrow\infty$. Dynamic heterogeneity is only a transient phenomenon. 

The time evolution of the van Hove function illuminates the behaviour found above for the self-intermediate function and the MSD for short particle displacements. 
Indeed, the coexistence of the two length scales $\ell_c$ and $\ell_a$ suggest that the non-trivial superdiffusion exponent at small $\Delta t$ in the MSD 
($\Delta r^2 \sim \Delta t^{1.4}$ in Fig.~\ref{fig:aging}(a)) 
is explained by the mixing of the two length scales: $\ell_c\propto\Delta t$ and $\ell_a\propto\Delta t^0$ at small $\Delta t$. 
Instead, the $q$-dependent relaxation time $\tau(q)$ in Fig.~\ref{fig:aging}(e) is mainly dominated by the behaviour of the length scale $\ell_c$, 
thus explaining its ballistic scaling,  $\tau(q)\propto q^{-1}\iff\ell_c\propto\Delta t$.
This follows from the fact that the measurement of $F_s(q,\Delta t)$ is dominated by the vast majority of the particles belonging to the core of the distribution (described by $\ell_c$), whereas the MSD is significantly influenced by the minority of fast moving particles (these particles make a large contribution to the MSD). 

\subsection{The key role of the global expansion}

\begin{figure}
\begin{centering}
\includegraphics[width=1.\columnwidth]{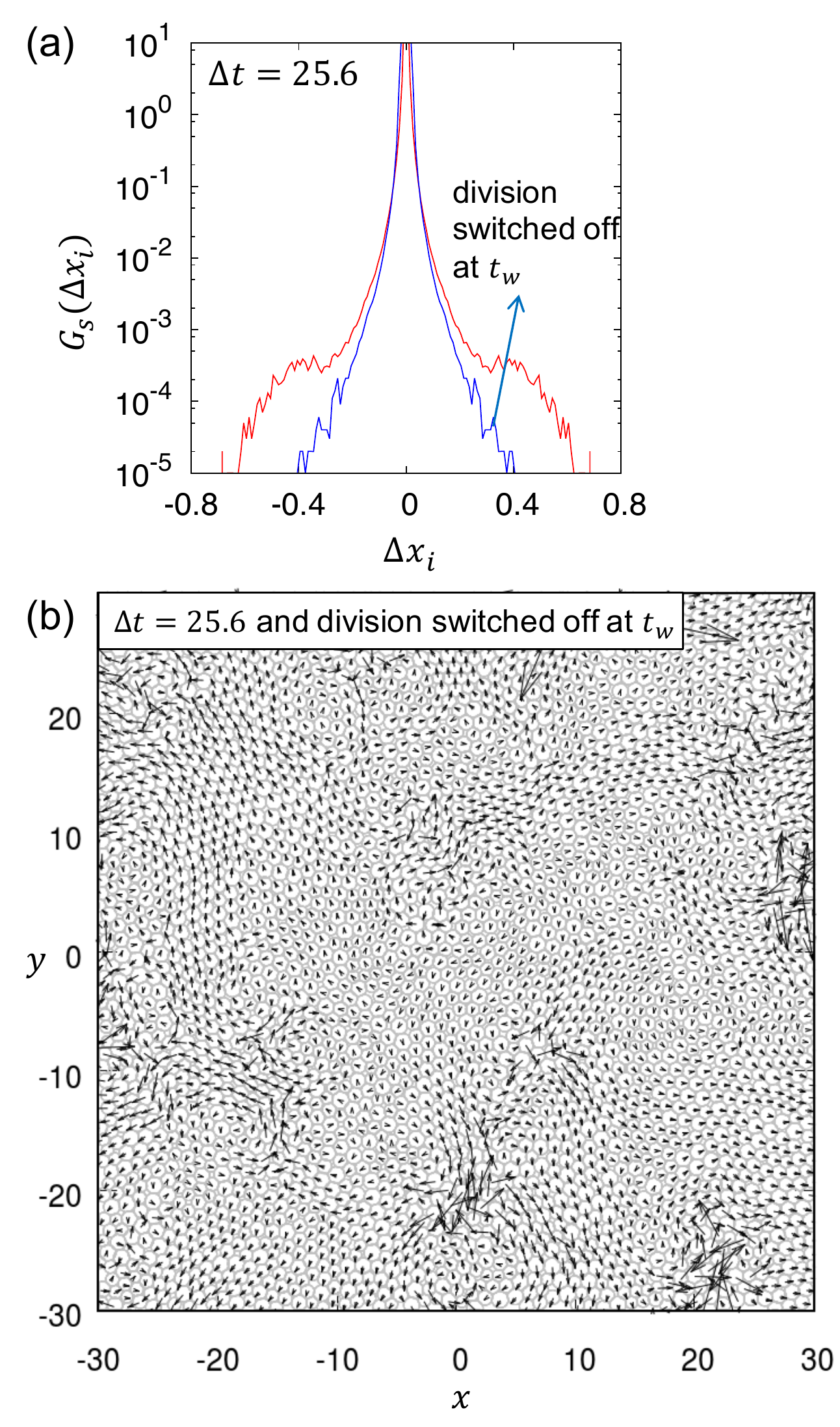}
\par\end{centering}
\caption{(a, blue) shows the same van Hove function as in Fig.~\ref{fig:heterogeneity}(a) (reproduced in red) but with division events completely switched off from time $t_w=104858.0$.
(b) Corresponding non-affine displacement field (again magnified by a factor 120), which is very similar to that of Fig.~\ref{fig:heterogeneity}(b), except for the division events. 
\label{fig:no-division}}
\end{figure}

We found that as the size of the growing system increases, the microscopic dynamics of the system slows down. One explanation for this could be related to the fact that as $t$ increases, the number of particles in the system increases linearly with time, which implies that the division rate per unit area also decreases as $t^{-1}$. Taking the view that division events are directly responsible for the fluidisation of the system would thus suggest that dynamics should indeed slow down with $t$. 

However, we find that the microscopic particle dynamics is actually not fully, or rather {\it not directly and not locally}, controlled by the division events. Instead, our interpretation is that the superposition of all division events in the system leads to a macroscopic expansion of the tissue, with a rate $\dot{\varepsilon}(t) \sim 1/t$ defined in Eq.~(\ref{eq:varepsilon}).  
 
To establish the key role played by $\dot{\varepsilon}$, we perform an independent simulation of our model where the dynamics proceeds normally up to a given time $t_w=104858.0$, after which we completely switch off all division events, letting the particle diameters grow with no bound (such simulation would of course become fully unphysical at large $\Delta t$ where particles would get very large).

The first effect of switching off division is to decrease the amplitude of the contribution of the fast moving particles in the van Hove function, as shown in Fig.~\ref{fig:no-division}(a), because division events no longer contribute to the tails, especially at very short times $\Delta t$. However, if we compare the displacement field in Fig.~\ref{fig:no-division}(b) and that in Fig.~\ref{fig:heterogeneity}(b), we see that the collective dynamics remains essentially the same, except for the few localised events indicated by red circles in Fig.~\ref{fig:heterogeneity}(b) which no longer appear in Fig.~\ref{fig:no-division}(b). However, several localised plastic events can still be observed. 
Therefore, our key point is that the non-affine displacement field remains spatially heterogeneous and highly complex even when division events are switched off. In addition, we find that the MSD remains essentially unaffected (i.e. superdiffusive) by the switching off of division events at short times, and is only reduced by small factor. 

These observations suggest that the collective and non-affine dynamics of the particles stem mostly from the fact that the colony is strained globally in the radial direction as a result of the macroscopic growth of the material. The rate of radial growth $\dot{\varepsilon}$ thus plays the role of a global macroscopic driving force on the dense amorphous colony, which then responds just as sheared glasses do. Namely, the forcing induces local plastic events (or shear transformation zones), which then lead to a stress redistribution at larger lengthscales. 

\section{Quasi-1D geometry}

\label{1D}

From the above, we showed that all dynamical properties of a radially growing dense active matter can be collapsed into a single universal function, 
which can be characterized by a single global parameter $\dot{\varepsilon}(t)$ and a universal exponent $\mu$. This is the same way as sheared soft glasses can be characterized by their strain rate.

\begin{figure}
\begin{centering}
\includegraphics[width=1.\columnwidth]{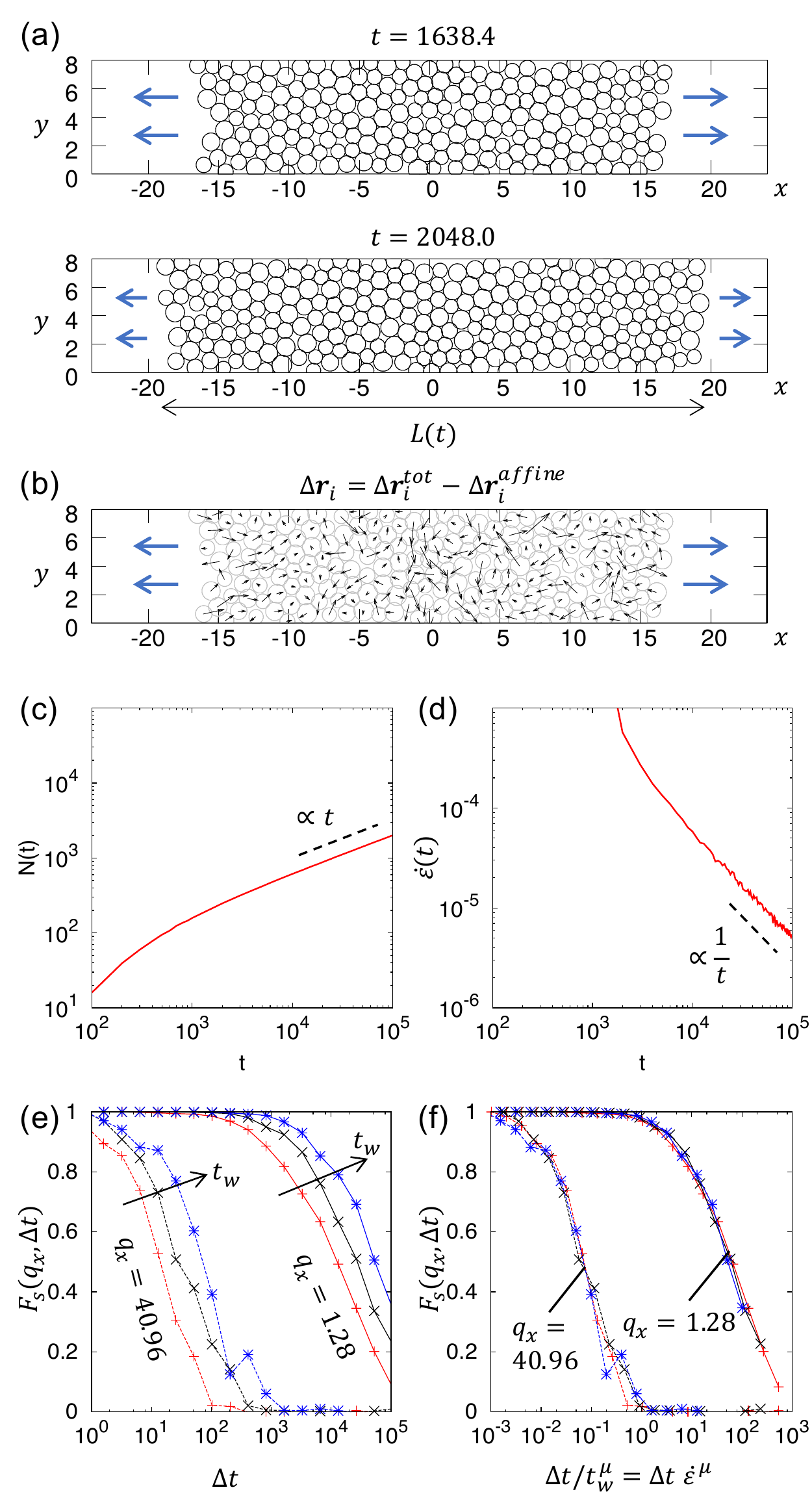}
\par\end{centering}
\caption{
(a) Snapshots of growing dense active matter in quasi-1D geometry. $L(t)$ is the lateral size of the colony at time $t$.
(b) shows the non-affine displacement field between time $t_w=1638.4$ and $t_w+\Delta t=2048.0$.
(c) The total number of particles increases linearly with time $t$ for large $t$.
(d) The strain/growth rate is defined to be $\dot{\varepsilon}(t)=\dot{L}/L$, which decreases as $1/t$.
(e,f) show the intermediate scattering function $F_s(q_x,t_w,\Delta t)$, which can be collapsed into a single universal function by rescaling the $x$-axis into $\Delta t/t_w^\mu$,
with $\mu\simeq0.6$. $q_x$ is the $x$-component of the wavevector.  
\label{fig:1D}}
\end{figure}

To test the generality of this result, we now consider a quasi one-dimensional (1D) geometry, where we assume periodic boundary condition at $y=0$ and $y=L_y$
and infinite domain in the $x$-direction (see Fig.~\ref{fig:1D}(a)).
This geometry is similar to experiments of wound healing in epithelial tissues~\cite{Martin11-wound-healing}.
We fix $L_y=8$ (in simulation units).
We initialize a strip of $8$ particles of unit diameter at $x=0$ at time $t=0$, and let the system evolve according to the equations of motion Eqs.~(\ref{eqmotion}-\ref{eq:rate}), as before.
Figure~\ref{fig:1D}(a) shows the snapshots of the dense active matter growing laterally in the $x$-direction at two different times.
We also define $L(t)$ to be the lateral size of the growing colony at time $t$ (analogous to $R(t)$ in the radial geometry considered above).
Similarly, we define the lateral growth rate to be $\dot{\varepsilon}(t)=\dot{L}/L$ in that case. 

For large enough time $t$, the total number of particles will increase linearly with time $t$ (see Fig.~\ref{fig:1D}(c)). 
This corresponds to the linear growth regime, equivalent to Fig.~\ref{fig:model}(d).
In this regime, the growth rate again decays as $\dot{\varepsilon}\sim1/t$ (see Fig.~\ref{fig:1D}(d)).

To make the same analysis as before, we must first subtract the affine components of the particles' displacements, 
which is the $x$-component of the integral in Eq.~(\ref{eq:aff}).
Figure~\ref{fig:1D}(b) shows the typical \emph{non-affine} displacement field of the particles between time $t_w$ and $t_w+\Delta t$.
We can then compute the same quantities such as the intermediate scattering function, defined in Eq.~(\ref{eq:Fs}),
except now, we only consider the $x$-component of the particles' displacements.
We find that the intermediate scattering function can also be collapsed into a near universal function by rescaling the delay time $\Delta t$ into $\Delta t\cdot\dot{\varepsilon}^\mu$ (see Fig.~\ref{fig:1D}(e,f)).
Thus the collapse hypothesis still works for different geometries, however, we find that the aging exponent is weakly dependent on the dimensionality: 
$\mu\simeq0.6$ in quasi-1D, compared to $\mu\simeq0.7$ in the radial geometry.
We also find a compressed exponential form of $F_s$, similar as before.

\section{Discussion and perspectives}

\label{conclusion}

\subsection{Physical discussion}

Thinking about crowded biological systems using analogies with dense glassy materials is a fruitful area~\cite{reviewelijah}. For instance, the competition between self-propulsion and particle crowding leads to the emergence of a non-equilibrium class of glass transitions~\cite{BB13}, with their own specific features~\cite{berthier14,elijah} 

The role of cell division has also been tackled in this context, with different views. Introducing both cell division and cell death can lead for instance to a non-equilibrium steady state where the number of cells can be  constant on average. This situation has been studied using several approaches~\cite{joanny,silke}, and it was concluded that this leads to visco-elastic behaviour, the short-time solid behaviour giving way to long-time flow controlled by the rate of cell division. 

Here, we studied a different case where particle division is not compensated by apoptosis, and thus the number of particles increases. 
This gives rise to the competition between steric repulsion and particle division again, but now inside a material that is macroscopically expanding, which we termed {\it growing active matter}. Although activity (and thus energy) is injected at the particle scale only by the division process, we found that resulting material expansion at the macroscopic scale gives rise to a finite expansion rate $\dot{\varepsilon}$ which is quite homogeneous inside the expanding system. As a result, the system becomes equivalent to a dense disordered assembly of particles submitted to a {\it global} mechanical forcing. Numerous studies of driven amorphous materials showed that this situation leads to plasticity and flow, taking the form of localised shear transformation zones which then redistributes the stress over large length scales, due to the elasticity. 

As a result, we found that the microscopic dynamics of the system is never arrested but instead shows relaxation over a time scale that is directly controlled by the radial growth rate, since the variable $\Delta t / t_w^\mu \approx \Delta t\cdot\dot{\varepsilon}^\mu$ essentially rescales all time correlation functions measured numerically. Interestingly, such non-linear rescaling with the strain rate $\dot{\varepsilon}$ has been reported in many types of sheared dense suspensions~\cite{Poon07,ballauff}. In the rheological context, dense suspensions are often sheared at constant shear rate $\dot{\gamma}$ in a simple shear geometry, but it is expected that the specific geometry of the external forcing is unimportant. 

Regarding the value of the exponent $\mu$, the simplest estimate should be $\mu = 1$ which would then correspond to the rheological behaviour of a simple yield stress material~\cite{review-yield}. Our finding $\mu \approx 0.7$ differs from this simple estimate. We see two possible explanations. 
First, even if the effective rate of deformation $\dot{\varepsilon}$ is the main control parameter, different rheological exponents with $\mu < 1$ are often observed in sheared glasses~\cite{review-yield}. Second, even though particle divisions is not uniquely driving the relaxation, these events occur and can speedup the relaxation by some amount, again reducing $\mu$ from unity.   

An experimental study of a soft gel sedimenting over gravity represents another relevant analogy~\cite{Cipelletti11}. In that case, the sedimenting gel of decreasing height $h(t)$ is compressed with a global compression rate $\dot{\varepsilon} = \dot{h(t)}/{h(t)}$, and it was also found that time correlation functions inside the material scale with the rescaled variable $\dot{\Delta t}/\dot{\varepsilon}^\mu$, with an exponent $\mu \approx 1.0$, with similar features as the ones reported above for the expanding active material. 

Our detailed analysis of particle motion at the particle scale clearly reveals the existence of localised plastic events (both due to particle division but also due to the plasticity induced by the global expansion of the material). The similarity with particle-resolved studies of sheared amorphous solids and computer simulations is again very striking, thus comforting our broad conclusion that dense growing active matter and sheared amorphous solids can be described by the same underlying physics.  

The analogy drawn here between an active system and a globally driven one differs qualitatively from a series of recent studies regarding the effect of self-propulsion on dense particle assemblies~\cite{BB13,berthier14,elijah}, which concluded that self-propulsion leads to glassy dynamics similar in many respects to thermally driven particle suspensions, so that the local driving force translates into a kind of non-driven, equilibrium-like relaxation dynamics. 

A third type of analogy was drawn from studying the effect of `self-deformation' (i.e. volume fluctuations occurring at the particle scale), where the activity was shown to give rise to local yielding events~\cite{Elsen17} that are then able to fluidise the system, but this fluidisation occurs without provoking a large-scale deformation as we found here.  

It is remarkable that these three types of analogies, derived from studies where some kind of ``activity'' is added to an otherwise densely packed disordered collection of soft objects (colloids, cells, bacteria), do not lead to a unique phenomenology, but instead to qualitatively different types of analogies with the behaviour of amorphous systems. We conclude that the physics observed in ``dense active matter'' may depend quite explicitely on the specific type of activity considered~\cite{reviewelijah}. 

\subsection{Perspectives}

In conclusion, we have studied a minimal model of growing dense active matter, a class which includes biological tissues, bacterial colonies and biofilms.

We have shown that the expanding nature of the material is key to understand its dynamics. In particular, it is useful to introduce the distinction between affine deformation due to the growth of the material, and the non-affine component which contains the relevant information of structural relaxation and microscopic dynamics. These dynamics shows pronounced dynamic heterogeneities, aging behaviour, and displays non-trivial time correlations functions exhibiting a crossover from ballistic motion of short scales accompanied by compressed exponential decay, crossing over to subdiffusive motion at larger scales.  

We have intentionally used a simplified model for expanding active matter which contains particle division and steric repulsion as unique ingredients to study their competition independently of any other physical processes that could of course be present in real materials, in particular biological ones. 
We are well aware that particle adhesion, self-propulsion, thermal fluctuations, internal structure or geometry of cells or bacteria may all affect the behaviour described in this work. We are currently exploring how to include some of these ingredients in more elaborated models, as well as collecting literature data for particle-resolved dynamic investigations of expanding biological systems. The study of the one-dimensional geometry in Sec.~\ref{1D} added for the revision of this manuscript is an effort in that direction. Our overall goal is to confirm that the main results of our study can be experimentally relevant. We also hope that our analysis, which borrows the tools developed to study dense amorphous solids, will motivate further experimental studies of the many fascinating examples of growing active matter that biology provides us.  

\begin{acknowledgments}
We thank M. Cates, N. M. Oliveira, and D. Thirumalai for useful exchanges about this work. This work was supported by a grant from the Simons Foundation (\#454933, L. Berthier). E. T. is funded in part by the European Research Council under the Horizon 2020 Programme, ERC Grant Agreement No. 740269.
\end{acknowledgments}

\end{document}